\documentclass[aps,prl,reprint,superscriptaddress,longbibliography]{revtex4-2}

\usepackage{amsmath,amssymb,bm}
\usepackage{graphicx}
\usepackage{booktabs}
\usepackage{xcolor}
\usepackage{hyperref}
\newcommand{\Tr}{\operatorname{Tr}}
\newcommand{\cN}{\mathcal{N}}
\newcommand{\cP}{\mathcal{P}}

\begin{document}

\title{Quantum Representation Learning Beyond Pairwise Fidelity}

\author{Junpeng Hou}
\email{jhou@pinterest.com}
\affiliation{Pinterest Inc., San Francisco, California 94103, USA}
\affiliation{Department of Physics, Washington University in St. Louis, St. Louis, MO 63130, USA}
\author{Changbin Lu}
\affiliation{School of Computer Science and Technology, Anhui University of Technology, Maanshan, China}

\begin{abstract}
Quantum contrastive, metric, and self-supervised learning often expose encoded
quantum states to the learner through transition probabilities,
especially fidelity. Quantum states are known to possess
higher-order relational invariants, but their consequences for learned
representations remain unclear. Here we show that a
transition-probability-only learning interface can possess exact continuous
blind directions in certain quantum-state families. We
recover this missing information with a batch operator, built from coherent overlap amplitudes and negative masking, where its second moment
$q_-$ retains four-state interference. Moreover, $q_-$ is directly measurable through two-copy interference and can
enter variational learning via methods like parameter shift. In relational
quartets derived from toric-code and double-semion states, this fidelity-blind
signal encodes inequivalent modular data despite identical pairwise
fidelities. Finally, in a four-photon benchmark
with preparation drift, augmenting all six pairwise fidelities at two
orthogonal probes with the corresponding normalized $q_-$ reduces
the mean out-of-distribution phase error by $86\%$ at equal total shot budget. These results establish multistate relational observables as measurable, trainable, and physically consequential signals for quantum representation learning beyond pairwise fidelity.
\end{abstract}

\maketitle

\textit{\textcolor{blue}{Introduction}.--}
Quantum representation learning (QRL) maps data into quantum states and uses
relations among those states as learning signals. In classical machine
learning, contrastive objectives organize representations by bringing related
examples together while separating unrelated ones
~\cite{Oord2018,Chen2020,Khosla2020,Radford2021}. Quantum embeddings and
feature-space methods provide a different representation geometry through
interference and entanglement~\cite{Havlicek2019,Cerezo2021}, motivating
quantum metric-learning approaches~\cite{Lloyd2020,LiuNQE2025} and, more
recently, quantum self-supervised and contrastive schemes based on
variational circuits and measured state overlaps
~\cite{Jaderberg2022,Wang2023,Chen2024,Don2025,Li2026,Zhukas2025}.
Across many such approaches, however, the relation presented to the learner
is reduced to a two-state scalar, with the transition probability $F_{ij}=|A_{ij}|^2, A_{ij}=\langle\psi_i|\phi_j\rangle$
as the canonical example.

Two-state transition probabilities are known not to determine, in general,
the relational geometry of quantum states. Gauge-invariant closed products of
overlaps and their associated Bargmann phases are established multistate
quantities that can contain information absent from transition probabilities
~\cite{Bargmann1964,Rabei1999,Mukunda2003,Oszmaniec2024,ChienWaldron2016}.
Such phase-sensitive relations have direct physical significance in settings
including topological order and multiparticle interference
~\cite{Kitaev2003,LevinWen2005,Smith2020Sign,Menssen2017,Jones2020}, while
multi-copy interferometry of nonlinear state functionals and
relational invariants is also well established
~\cite{Ekert2002,Islam2015,Oszmaniec2024}. Recent protocols further target
efficient nonlinear-property estimation under replica or purified access
~\cite{LiuReplica2026,ZhangNonlinear2026}. The unresolved issue for
QRL is then what is lost when a learner is given only transition
probabilities? Can such an interface possess blind
directions in the representation space? If so, can the missing information be made measurable, trainable,
and useful?

\begin{figure}[t]
    \centering
    \includegraphics[width=\columnwidth]{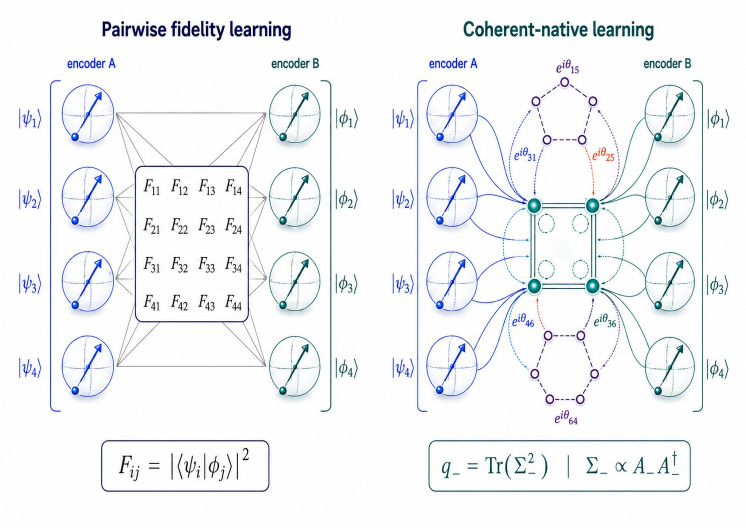}
    \caption{
    \textbf{From transition probabilities to multistate relational learning.}
    \emph{Left:} A common QRL interface reduces relations to two-state
    quantities such as fidelity $F_{ij}$.
    \emph{Right:} Coherent access to overlap amplitudes preserves multistate
    interference, with $q_-$ providing a nonlinear spectral observable of
    the masked relational structure.
    }
    \label{fig:concept}
\end{figure}

Here we show that this blindness can be exact and introduce a learning
interface that restores the missing information, as shown in Fig.~\ref{fig:concept}. We construct $\Sigma_-\propto A_-A_-^\dagger$ from the coherent overlap amplitudes selected by a learning
mask. Its trace reproduces the mean negative
fidelity, whereas the second moment
$q_-=\Tr\Sigma_-^2$ is a mask-conditioned aggregate of four-state
interference loops.
After conditioning on the overlap branch, the normalized $\widetilde q_-$ is exactly the purity of the resulting bipartite relational
state. Thus, it admits two complementary physical readings:
spectrally, it measures the concentration of overlap weight into collective
relational modes; moreover, it measures coherent interference
around closed relational loops.
Using a continuous complex-Hadamard family,
we exhibit a relational coordinate for which every two-state
transition probability is fixed while $q_-$ varies continuously.
Consequently, any differentiable objective whose relational inputs consist
only of these transition probabilities is exactly flat along this direction.
Meanwhile, the multistate observables can be obtained directly from
two-copy interference and they can admit
parameter-shift gradients for
variational learning.

We establish three cascading experiments. First, in a controlled
two-qubit realization, contrastive optimization selects complementary
quadratures within a parameterized coherent-probe family, demonstrating the
trainability. Second, relational quartets
derived from microscopic toric-code and double-semion states have identical
complete pairwise fidelities but different coherent moments, showing that fidelity-blind relational information can encode
inequivalent modular data. Finally, in a physics-grounded simulated
four-photon interference benchmark, positive views share a collective phase
while undergoing independent preparation drift. At a fixed two-probe
measurement design, a label-free learner given only fidelities
always fails with out of distribution (OOD) tests, whereas adding the normalized moments restores phase generalization.
These results motivate a broader view of QRL in which quantum mechanics enters not only through the states, but also through the learning objectives via multistate relational information.

\textit{\textcolor{blue}{Coherent relational observables}.--}
Consider two parameterized quantum encoders producing normalized states
$|\psi_i(\bm\theta_I)\rangle$ and
$|\phi_j(\bm\theta_T)\rangle$. Given labeled data, let
$\cP=\{(i,i)\}_{i=1}^{B}$ and
$\cN=\{(i,j):i\neq j\}$ denote positive and negative pairs, with $M=|\cN|=B(B-1)$ and $B$ the batch size.
Assuming state-preparation unitaries
$U_i^I|0^n\rangle=|\psi_i\rangle$ and
$U_j^T|0^n\rangle=|\phi_j\rangle$, a uniform superposition over $\cN$ followed by the address-controlled overlap operation prepares $|\Psi_-\rangle=
\frac{1}{\sqrt M}
\sum_{(i,j)\in\cN}
|i,j\rangle\,
(U_i^I)^\dagger U_j^T|0^n\rangle$.
Projecting the work register onto $|0^n\rangle$ selects the
address state $|g_-\rangle=
\frac{1}{\sqrt M}
\sum_{(i,j)\in\cN}
A_{ij}|i,j\rangle$.
Tracing out one address register defines the positive batch-level, generally subnormalized relational
operator $\Sigma_-=\Tr_T|g_-\rangle\langle g_-|=\frac{1}{M}A_-A_-^\dagger$,
where $A_-$ absorbs the learning or negative mask $\cN$.

Its first two spectral moments separate the total pairwise-overlap weight from
its coherent organization,
\begin{equation}
p_-=\Tr\Sigma_-=
\frac{1}{M}\sum_{(i,j)\in\cN}|A_{ij}|^2,~
q_-=\Tr\Sigma_-^2 .
\label{eq:qminus}
\end{equation}
Thus $p_-$ is exactly the mean negative fidelity.
Since $\langle g_-|g_-\rangle=p_-$, conditioning on the overlap branch defines the
normalized bipartite relational state
$|\widehat g_-\rangle=|g_-\rangle/\sqrt{p_-}$, whose reduced state is
$
\rho_-=
\Tr_T|\widehat g_-\rangle\langle\widehat g_-|
=
\frac{\Sigma_-}{p_-}
$.
The normalized second moment is therefore its purity,
$\widetilde q_-=\Tr\rho_-^2=q_-/p_-^2$. At fixed $p_-$,
$\widetilde q_-$ measures how the relational weight is distributed among
Schmidt modes: it equals unity for a single mode and decreases as the weight
spreads over orthogonal modes. Thus
$q_-=p_-^2\widetilde q_-$ combines the total pairwise-overlap weight with its
coherent relational organization.

Upon expanding $q_-$, its phase-sensitive contributions contain four-state
gauge-invariant Bargmann-type products
$B_{ikjl}=\langle\psi_i|\phi_j\rangle
\langle\phi_j|\psi_k\rangle
\langle\psi_k|\phi_l\rangle
\langle\phi_l|\psi_i\rangle$
\cite{Bargmann1964,Rabei1999,Mukunda2003,Oszmaniec2024}.
Writing $A_{ij}=\sqrt{F_{ij}}e^{i\theta_{ij}}$, their interference phase is
$\theta_{ij}-\theta_{kj}+\theta_{kl}-\theta_{il}$, which is invariant under
independent rephasings of the individual states. Pairwise fidelities therefore
specify the edge intensities of the relational graph, whereas $q_-$ also
detects interference between distinct coherent paths around closed
four-cycles. The learning mask determines which such loops contribute. For
the standard mask $i\neq j$, the first phase-sensitive loop appears at
$B=4$ and $q_-$ is the lowest spectral moment that closes a
gauge-invariant interference loop.

This leads to an exact information separation. Consider the complex-Hadamard family $H_4(\varphi)=\frac12
\begin{pmatrix}
1&1&1&1\\
1&e^{i\varphi}&-1&-e^{i\varphi}\\
1&-1&1&-1\\
1&-e^{i\varphi}&-1&e^{i\varphi}
\end{pmatrix}$ ~\cite{Tadej2006}.
Let $\{|\psi_i\rangle\}$ be the computational basis and
$\{|\phi_j(\varphi)\rangle\}$ the columns of $H_4(\varphi)$.
Both sets are orthonormal, while $|\langle\psi_i|\phi_j(\varphi)\rangle|^2=\frac14$
for every $i,j$ and $\varphi$. Therefore all two-state transition
probabilities are constant.
The masked coherent second moment, however, varies continuously as $q_H(\varphi)\propto\cos\varphi$.

Let $\mathcal L_{\rm TP}$ denote any differentiable learning objective whose
relational inputs consist only of two-state transition
probabilities. Given the $H_4(\varphi)$ family, we have
\begin{equation}
\frac{d\mathcal L_{\rm TP}}{d\varphi}=0,\quad
\frac{dq_H}{d\varphi}
=
-\frac{1}{144}\sin\varphi .
\label{eq:blind-direction}
\end{equation}
Thus $\varphi$ is an exact continuous blind direction of the
transition-probability-only learning interface.

\textit{\textcolor{blue}{Direct measurement and trainability}.--}
The relational moment does not require tomography or classical reconstruction
of the overlap matrix. Under coherent indexed access,
the projection probability of the work register onto $|0^n\rangle$ gives
$p_-$. For two copies, define
$\Pi_0=\mathbb{I}_{IT}\otimes|0^n\rangle\langle0^n|_W$ and let $S_I$ exchange only
the $I$-address registers. The standard two-copy identity gives
\begin{equation}
q_-=
\langle\Psi_-|^{\otimes2}
\Pi_0^{(1)}\Pi_0^{(2)}
S_I
|\Psi_-\rangle^{\otimes2},
\label{eq:q-circuit}
\end{equation}
which is exactly
$\Tr[(\Tr_T|g_-\rangle\langle g_-|)^2]$.
Since the probability that each copy occupies the good branch is $p_-$,
conditioning on both good branches gives
$q_-/p_-^2=\Tr\rho_-^2$. The same two-copy measurement therefore has a
direct purity interpretation: the raw estimator yields the overlap-weighted
relational purity $q_-$, while conditioning removes the total overlap scale
and yields the relational purity itself. Thus standard nonlinear multi-copy
measurements~\cite{Ekert2002,Islam2015,Oszmaniec2024} become, after the masked
coherent overlap construction, direct estimators of a batch-level relational
learning observable.

\begin{figure}[t]
    \centering
    \includegraphics[width=\columnwidth]{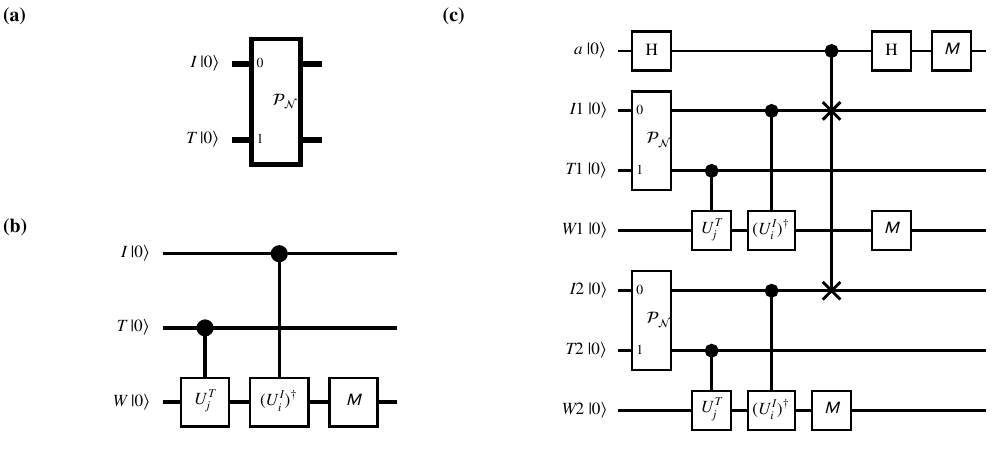}
    \caption{
    \textbf{Direct access to multistate relational observables.}
    (a) A coherent loader prepares the desired pair superposition.
    (b) The address-controlled overlap operation writes
    $A_{ij}$ into the $|0^n\rangle_W$ branch, whose probability gives $p_-$.
    (c) For two copies, projection followed by a SWAP
    of one address subsystem yields $q_-=\Tr\Sigma_-^2$.
    }
    \label{fig:circuit}
\end{figure}

The controlled-SWAP in Fig.~\ref{fig:circuit} gives one implementation: a destructive Bell-basis readout measures the desired observable. Because the encoded states are parameterized, $q_-$ also
admits parameter-shift gradients and can therefore enter variational
optimization directly. Finite-shot estimators, destructive-SWAP circuits,
gradient identities, and logical resource counts are given in the
Supplemental Material.

Having established both an exact fidelity-blind direction and a corresponding
trainable observable, we next ask the central learning
question: can multistate relational information form a useful
representation?

\textit{\textcolor{blue}{Learning the relational measurement}.--}
In the exact separation $q_H(\varphi)$ shown above, $\varphi$ continuously changes the coherent organization of the
relational state while every pairwise transition probability remains fixed.
A single coherent setting nevertheless cannot distinguish $\pm\varphi$ since
$q_H(\varphi)=q_H(-\varphi)$. We therefore introduce a normalized pair
$\bigl(\cos\varphi,\cos(\varphi+\delta)\bigr)$ with a second trainable phase
offset $\delta$.
Positive views are
independently perturbed realizations of the same geometry, and the phase
labels are never supplied during training. Complementary information is
obtained at $\delta=\pm\pi/2$.

\begin{figure}[t]
\centering
\includegraphics[width=\columnwidth]{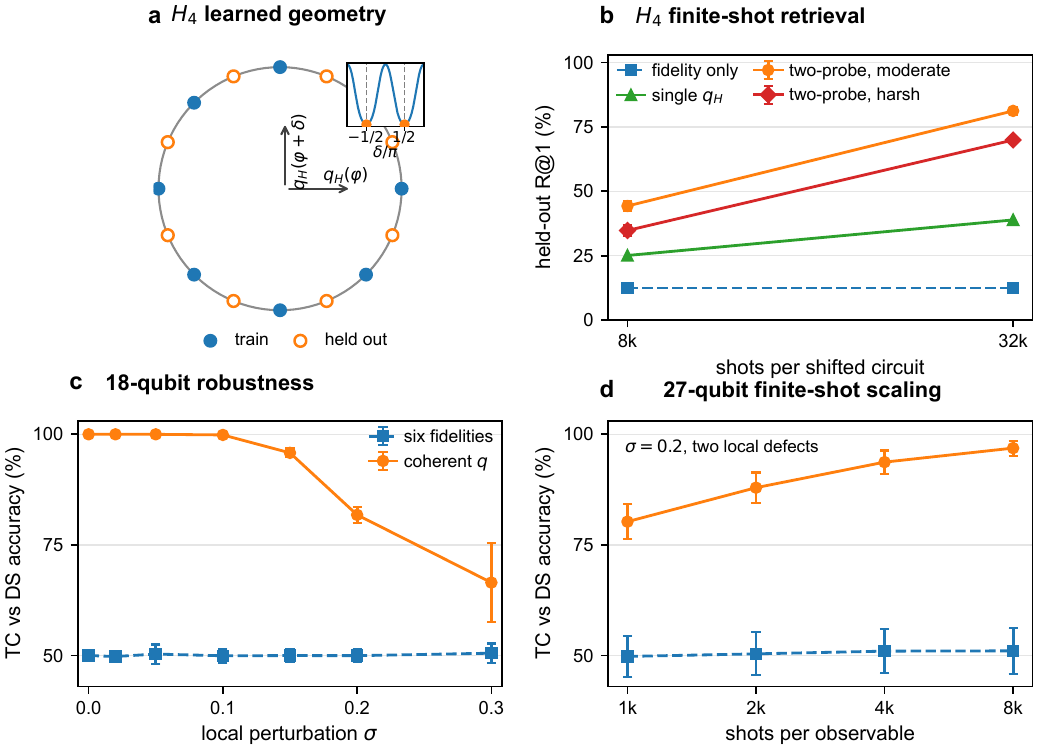}
\caption{
\textbf{Learning and physical content of multistate relations.}
(a) Contrastive learning of the complex-Hadamard family. Inset shows the
contrastive landscape.
(b) Recall on held-out retrieval under finite-shot readout.
(c) Microscopic TC-DS identification
for 18-qubit loop-gas states under independent local $R_x$ perturbations.
(d) Finite-shot topological identification for 27-qubit microscopic states
with two local defects per prepared ray. Pairwise fidelities remain
uninformative, while $q_-$ retains the topological distinction.
}
\label{fig:learning}
\end{figure}

Contrastive optimization with alternating train / heldout data points selects precisely the quadrature solutions
(see Fig.~\ref{fig:learning}(a)), yielding a representation equivalent to
$(\cos\varphi,\sin\varphi)$ and recovering the latent circle. Thus the learner is able to select
complementary quadratures within a prescribed coherent-probe family.
The selection remains robust to finite sampling: under the moderate noise
model, 29 of 30 optimizations at 8192 shots converge within $0.25$ rad of a
quadrature solution (Fig.~\ref{fig:learning}(b)). Full circuit,
optimization, retrieval, and noise details are given in the Supplemental
Material.

\textit{\textcolor{blue}{Topological information beyond pairwise relations}.--}
The fidelity-blind information is not restricted to a synthetic
geometry. We consider fixed-point toric-code (TC) and double-semion (DS)
orders on periodic honeycomb lattices~\cite{Kitaev2003,LevinWen2005,Zhang2020Signs}.
With qubits on edges, their ground states in homology sector $w$ have the
loop-gas form
\begin{equation}
|\Omega_w^{\rm DS}\rangle
=
\frac{1}{\sqrt{N_w}}
\sum_{c\in\mathcal C_w}
(-1)^{N_\ell(c)}|c\rangle ,
\label{eq:topo-loopgas}
\end{equation}
where $N_\ell(c)$ counts closed-loop components. From their microscopic
ground spaces we form a four-state relational quartet using the modular
transformations $S$ and $T$, which encode the distinct anyonic statistics of
the two orders~\cite{MoradiWen2015,LiMong2022}. The explicit construction is
given in the Supplemental Material.

Remarkably, the complete pairwise fidelities of the two quartets are
identical $\{F_{ab}\}=
\left\{
\frac14,0,\frac14,\frac14,\frac14,\frac14
\right\}$,
and their total overlap weights are also equal. The multistate relations
nevertheless remain distinct:
$q_{\rm TC}=\frac{1}{16}$,
$q_{\rm DS}=\frac{3}{64}$,
$\arg\mathcal B_{\rm TC}=0$, and
$\arg\mathcal B_{\rm DS}=\frac{\pi}{2}$.
Since both quartets have $p=\frac14$, their normalized relational purities are
$\widetilde q_{\rm TC}=1$ and $\widetilde q_{\rm DS}=\frac34$.
Thus the transition-probability interface erases modular-phase information
that reappears as a change in the purity and coherent mode structure of the
relational state. This result provides an exact fidelity-blind encoding of
inequivalent topological order.

The distinction survives microscopic perturbations and finite sampling.
After lifting the quartet into explicit loop-gas wavefunctions and applying
independent local $R_x$ perturbations to physical edge qubits, all
pairwise fidelities remain noninformative while $q$ retains high accuracy (Fig.~\ref{fig:learning}(c)). The effect persists with increasing
system size with a 27-qubit realization: a readout based on $q$ reaches up to $96.9\pm1.7\%$ separation accuracy, whereas the fidelity baseline remains at chance
(Fig.~\ref{fig:learning}(d)). Full perturbation, size-scaling, and finite-shot
results are given in the Supplemental Material.

The microscopic TC and DS states and
their local perturbations are constructed explicitly, but the modular
transformations defining the quartet are supplied as exact logical operations
on the ground space. Moreover, MES and ground-state overlaps can carry labeling and phase
ambiguities in general~\cite{LiMong2022}. Thus, this only confirms
fidelity-blind separation of relational quartets derived from inequivalent
modular data, but does not support an unsupervised discovery of a modular protocol.

\textit{\textcolor{blue}{Collective-phase learning and generalization}.--}
A significant consequence of retaining multistate relational information is that it enhances generalization under physical distribution shift.
We show this in multiphoton interference, where collective
phases are known to carry information beyond pairwise distinguishability
~\cite{Menssen2017,Jones2020,Rodari2026}. Specifically, we construct a
physics-grounded simulation parameterized by a four-photon experiment~\cite{Jones2020}. Each example is labeled physically by a continuously varying
collective phase $\Phi$, while temporal overlap, residual distinguishability,
polarization balance, and spectral mismatch act as preparation nuisances.

Two independently drifted realizations of the same $\Phi$ form a positive
pair, and $\Phi$ itself is never supplied during learning.
The strongest transition-probability baseline receives all
pairwise fidelities at each of two phase probes separated by $\pi/2$.
The coherent representation receives exactly the same fidelities and adds
only the corresponding normalized second moments
$\widetilde q_-$, namely the purities of the
postselected relational states. This normalization removes fluctuations in
the total overlap scale while retaining phase-sensitive organization of the
relational modes.
The two purity probes therefore expose complementary quadratures of the
collective phase. We use tied linear canonical correlation analysis (CCA) so that the comparison only probes the information
made available by the two relational interfaces (not model capacity or optimizer). Only after the encoder is frozen are 16
calibration phases used to define a linear phase readout. Evaluation is on
interleaved unseen phases under a simultaneous shift of all nuisances.

\begin{figure}[t]
    \centering
    \includegraphics[width=\columnwidth]{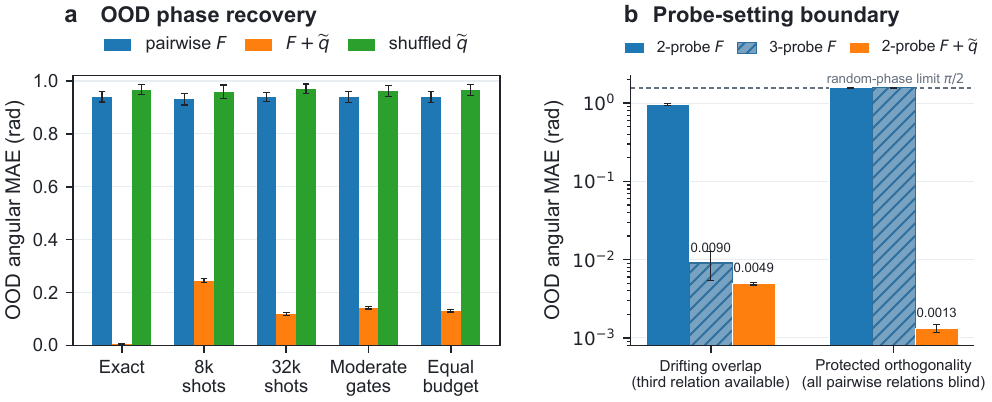}
    \caption{
    \textbf{Out-of-distribution learning and generalization.}
    All error bars denote the standard deviation over ten random seeds.
    (a) OOD angular mean absolute error of tied CCA using the complete
    two-probe pairwise-fidelity representation $F$, the representation regularized by normalized second moment $F+\widetilde q_-$, and a control in which
    $\widetilde q_-$ is shuffled across examples. Results are shown for exact
    observables, finite-shot sampling, the moderate gate-error model, and an
    equal-total-shot comparison.
    (b) Exact-expectation probe-number boundary. For generic drifting overlap,
    a balanced third fidelity probe restores identifiability. When the relevant
    temporal modes are exactly orthogonal, however, all phase-shifted pairwise
    fidelities remain blind and the pairwise errors stay at the random-phase
    value $\pi/2$, while the two-probe coherent representation remains
    informative.
    }
    \label{fig:photonic}
\end{figure}

The primary result is obtained at equal total measurement budget. A total of
$393{,}216$ shots is evenly distributed over 12/14 observables for the pairwise/coherent representation. The mean
OOD angular error reduces by 86\% from
$0.939\pm0.022$ rad to $0.131\pm0.005$ rad (Fig.~\ref{fig:photonic}(a)). The separation is already present at exact
expectation values, and survives explicit moderate gate and readout noises.
Thus simply increasing the measurement precision of a complete two-probe fidelity
description does not recover the information retained by $q_-$.

Two controls isolate the origin of the improvement. First, jointly shuffling
the two $\widetilde q_-$ coordinates across examples preserves their
feature statistics but destroys their association with the
underlying physical realization. At equal total shot budget, the OOD error
returns to $0.966\pm0.021$ rad
(Fig.~\ref{fig:photonic}(a)). The gain therefore does not arise merely from
supplying two additional features. Second, when the residual-overlap nuisance
that leaks phase information into a pairwise fidelity is held fixed, the
ambiguity disappears: an independent supervised probe reaches
$0.0266\pm0.0013$ rad using only the two-probe fidelities, compared with
$0.0191\pm0.0011$ rad after adding $\widetilde q_-$
(see Supplemental Material). Pairwise fidelity is not
intrinsically incapable of representing the collective phase.

The number and structure of active measurement settings provide a second
boundary. In the drifting-overlap model, three balanced fidelity probes reduce
the exact-expectation error to
$0.0090\pm0.0037$ rad (Fig.~\ref{fig:photonic}(b)). By contrast, setting the overlap of the relevant temporal
modes exactly to zero makes every phase-shifted pairwise fidelity independent
of the collective phase, which can occur for orthogonal time or frequency bins or
a symmetry-protected mode relation. In this protected limit, two- and
three-probe fidelity learners remain at $\pi/2$ error, whereas the two-probe
coherent representation reaches $0.00132\pm0.00016$ rad
(Fig.~\ref{fig:photonic}(b)). Thus, the advantage of multistate regularizer also depends on the physical system and experiment setup. Complete finite-shot and equal-budget probe
ablations are given in the Supplemental Material.

\textit{\textcolor{blue}{Conclusion and discussion}.--}
We have shown that transition-probability-only learning interfaces can possess
exact continuous blind directions, and that coherent multistate observables
can restore the missing information as measurable and trainable learning
signals. The same fidelity-blind structure can encode inequivalent
topological data and determine whether a learned representation remains informative OOD.

These results establish relational observables as a distinct degree of
freedom in QRL design. Beyond $q_-$, higher spectral moments and selected
interference loops provide natural extensions. Their utility must be weighed
against stronger access requirements, copy complexity, circuit depth,
sampling, and noise. It remains as a future study topic to understand
end-to-end computational costs and compare with classical routines. More broadly, quantum representations are determined not
only by the states produced by quantum encoders, but also by which collective
relations the learning interface preserves.

\begin{acknowledgments}
\textit{\textcolor{blue}{Acknowledgments}.--}
J. Hou thanks C. Zhang, J. Yan, and X. Lin for helpful discussions on quantum machine learning; A. Dharawat, J. Qu and Y. Chen for inspiring discussion about classical contrastive learning and negative mining. C. Lu acknowledges support by National Natural Science Foundation of China (Grant number 62501012) and Anhui Province University Natural Science Research Project (Grant number 2023AH051102).
\end{acknowledgments}

\makeatletter
\newenvironment{suppbibliography}[1]{%
 \def\NAT@bibsetnum##1{%
  \setlength{\topsep}{\z@}%
  \NATx@bibsetnum{\ref{LastSuppBibItem}}%
 }%
 \NAT@thebibliography{#1}%
}{%
 \edef\@currentlabel{\arabic{NAT@ctr}}%
 \label{LastSuppBibItem}%
 \endNAT@thebibliography
}%
\makeatother

\clearpage
\onecolumngrid
\setcounter{equation}{0}
\renewcommand{\theequation}{S\arabic{equation}}
\renewcommand{\theHequation}{S\arabic{equation}}
\setcounter{figure}{0}
\renewcommand{\thefigure}{S\arabic{figure}}
\renewcommand{\theHfigure}{S\arabic{figure}}
\setcounter{table}{0}
\renewcommand{\thetable}{S\arabic{table}}
\renewcommand{\theHtable}{S\arabic{table}}

\begin{center}
{\large\bf Supplemental Material for
``Quantum Representation Learning Beyond Pairwise Fidelity''}
\end{center}

\section{Overlap state and confusion operator}

\subsection{Coherent second moment}
Let $U_i^I|0^n\rangle=|\psi_i\rangle$ and $ U_j^T|0^n\rangle=|\phi_j\rangle$ (I for image and T for text as in conventional multi-modal contrastive learning setup).
For a negative set $\mathcal N$ of size $M$, coherent indexed access prepares
\begin{equation}
|\Psi_-\rangle=
\frac{1}{\sqrt M}
\sum_{(i,j)\in\mathcal N}
|i,j\rangle
(U_i^I)^\dagger U_j^T|0^n\rangle.
\label{eq:supp-psi}
\end{equation}
Expanding the work-register state into the component parallel to
$|0^n\rangle$ and its orthogonal complement,
\((U_i^I)^\dagger U_j^T|0^n\rangle = A_{ij}|0^n\rangle + |\chi_{ij}^{\perp}\rangle,  A_{ij}\equiv\langle\psi_i|\phi_j\rangle,\)
where
$\langle0^n|\chi_{ij}^{\perp}\rangle=0$.
The good component is
\(|g_-\rangle= \frac{1}{\sqrt M} \sum_{(i,j)\in\mathcal N} A_{ij}|i,j\rangle.\)

Tracing the second address register gives
\begin{align}
\Sigma_-
&=
\Tr_T|g_-\rangle\langle g_-|\\
&=
\frac{1}{M}
\sum_{i,k}
\left(
\sum_j A_{ij}A_{kj}^*
\right)
|i\rangle\langle k|\\
&=
\frac{1}{M}A_-A_-^\dagger,
\label{eq:supp-sigma}
\end{align}
where the mask is absorbed into $A_-$ so that $(A_-)_{ij}=A_{ij}$ on $\cN$ and vanishes otherwise.  The operator is positive
semidefinite but generally subnormalized.

Its first two moments are
\begin{align}
p_-&=\Tr\Sigma_-
=
\frac{1}{M}
\sum_{(i,j)\in\mathcal N}|A_{ij}|^2,
\label{eq:supp-p}\\
q_-&=\Tr\Sigma_-^2
=
\frac{1}{M^2}
\sum_{i,k,j,l}
A_{ij}A_{kj}^*A_{kl}A_{il}^*,
\label{eq:supp-q}
\end{align}
with the mask restrictions implicit in Eq.~\eqref{eq:supp-q}.

If $\lambda_r\ge0$ are the eigenvalues of $\Sigma_-$,
$p_-=\sum_r\lambda_r,
q_-=\sum_r\lambda_r^2.$
At fixed total negative weight, a larger $q_-$ therefore signals greater concentration into collective confusion modes. Useful consistency bounds are
\(0\le q_-\le p_-^2\le p_-\le1.\)
For fixed $p_-$, $q_-=p_-^2$ for rank-one confusion, whereas
$q_-=p_-^2/r$ when the weight is uniformly distributed over $r$
orthogonal modes.

\subsection{Gauge invariance and four-state interference loops}

Under independent rephasings,
\(|\psi_i\rangle\rightarrow e^{i\alpha_i}|\psi_i\rangle, |\phi_j\rangle\rightarrow e^{i\beta_j}|\phi_j\rangle,\)
the overlap matrix transforms as
$A_-\rightarrow D_I^\dagger A_-D_T,$
where $D_I$ and $D_T$ are diagonal unitaries. Hence
$\Sigma_-\rightarrow D_I^\dagger\Sigma_-D_I,$
and every spectral invariant $\Tr\Sigma_-^m$ is gauge invariant.

The phase-sensitive terms entering $q_-$ have the form
\begin{equation}
\mathcal B_{ikjl}
=
\langle\psi_i|\phi_j\rangle
\langle\phi_j|\psi_k\rangle
\langle\psi_k|\phi_l\rangle
\langle\phi_l|\psi_i\rangle.
\label{eq:supp-bargmann}
\end{equation}
They are four-state Bargmann-type invariants
~\cite{S-Bargmann1964,S-Rabei1999,S-Mukunda2003,S-Oszmaniec2024}.
Their phases are relational rather than arbitrary phases assigned to
individual state vectors.

The independence of multistate relational information from two-state data
depends on Hilbert-space dimension. In particular, single-qubit multistates
obey additional constraints on their higher-order Bargmann invariants
~\cite{S-LiWagnerZhang2026}. The explicit information-separation constructions
used here live in a four-dimensional Hilbert space and are realized with two
qubits. We do not claim that four dimensions are minimal and the
single-qubit case is a constrained special case, which is not used
for the separation established below.

\section{Fidelity-only information separation}

\subsection{Interference-loop criterion and the minimal paired batch}

For a general negative mask define
$\mathcal N(i)=\{j:(i,j)\in\mathcal N\}.$
Starting from Eq.~\eqref{eq:supp-sigma},
\begin{align}
M^2q_-
&=
\Tr[(A_-A_-^\dagger)^2]\\
&=
\sum_{i,k}
|(A_-A_-^\dagger)_{ik}|^2\\
&=
\sum_{i,k}
\left|
\sum_{j\in\mathcal N(i)\cap\mathcal N(k)}
A_{ij}A_{kj}^*
\right|^2.
\label{eq:supp-common-neighbor}
\end{align}

For $i=k$, the inner sum contains only the edge magnitudes
$|A_{ij}|^2$. For $i\neq k$, phase sensitivity requires interference
between two distinct common intermediate states $j\neq l$, giving
\begin{equation}
2\operatorname{Re}
\left[
A_{ij}A_{kj}^*A_{kl}A_{il}^*
\right].
\label{eq:supp-loopterm}
\end{equation}
The four required edges
$(i,j),\quad(k,j),\quad(k,l),\quad(i,l)$
form a length-four cycle in the bipartite mask graph. Physically, this
corresponds to two distinct coherent paths connecting the same pair of
learned states,
\(|\psi_i\rangle\rightarrow|\phi_j\rangle\rightarrow|\psi_k\rangle,  |\psi_i\rangle\rightarrow|\phi_l\rangle\rightarrow|\psi_k\rangle,\)
ans Eq.~\eqref{eq:supp-loopterm} is their interference term. Consequently, if the bipartite mask graph contains no length-four cycle,
$q_-$ is fixed by the edge fidelities $|A_{ij}|^2$. Cycle-based
organization of projective invariants is known in finite-frame geometry
~\cite{S-ChienWaldron2016}. The role of the learning mask here is to determine
which interference loops can enter the objective.

For the standard paired mask $\mathcal N=\{(i,j):i\neq j\}$,
two distinct image vertices share exactly $B-2$ negative text neighbors.
Two distinct coherent paths require $B-2\ge2$, so
$B_{\min}=4$
is the smallest paired batch for which $q_-$ can contain phase information
not determined by pairwise fidelities.

Independent rephasings transform the overlap matrix as
$A\rightarrow D_I^\dagger A D_T.$
A common unitary acting on both state sets leaves $A$ unchanged. A common
antiunitary maps $A$ to $A^*$, up to the same row and column phase freedom.
In all cases the singular values of $A_-$, and hence
$q_-=
M^{-2}\Tr[(A_-A_-^\dagger)^2],$
are unchanged. Ensembles with different $q_-$ therefore cannot be related
by these transformations. The separation is a property of their relational
geometry

\subsection{Complex-Hadamard exact separation}

A stronger geometric separation follows from the one-parameter order-four
complex-Hadamard family~\cite{S-Tadej2006},
\begin{equation}
H_4(\varphi)=\frac12
\begin{pmatrix}
1&1&1&1\\
1&e^{i\varphi}&-1&-e^{i\varphi}\\
1&-1&1&-1\\
1&-e^{i\varphi}&-1&e^{i\varphi}
\end{pmatrix}.
\label{eq:supp-H4}
\end{equation}
Let $\{|\psi_i\rangle\}$ be the computational basis and
$\{|\phi_j(\varphi)\rangle\}$ the columns of $H_4(\varphi)$. Both sets are
orthonormal, and
$|\langle\psi_i|\phi_j(\varphi)\rangle|^2=\frac14$ remains a constant.

Therefore \emph{all} two-state transition probabilities among the eight
states are independent of $\varphi$: within either basis they are $0$ or
$1$, and across the two bases they are $1/4$. Nevertheless, after applying
the same paired mask $i\neq j$,
\begin{equation}
q_-^{H}(\varphi)
=
\frac{17+4\cos\varphi}{576}.
\label{eq:supp-qH4}
\end{equation}
This construction isolates the incompleteness of two-state transition
probabilities themselves and forms the quantum-native learning task used in
the main text.

\section{Measurement and training of the coherent second moment}

\subsection{Controlled-SWAP identity}

Define the good-space projector on one copy as
$\Pi_0
=
\mathbb{I}_{IT}\otimes|0^n\rangle\langle0^n|_W.$
For two copies of $|\Psi_-\rangle$, let $S_I$ swap only the image-address
registers. Since
$\Pi_0|\Psi_-\rangle
=
|g_-\rangle|0^n\rangle_W,$
we obtain
\begin{align}
&
\langle\Psi_-|^{\otimes2}
\Pi_0^{(1)}\Pi_0^{(2)}
S_I
|\Psi_-\rangle^{\otimes2}
\nonumber\\
=&
\langle g_-|^{\otimes2}
S_I
|g_-\rangle^{\otimes2}
\nonumber\\
=&
\Tr\left[
\left(
\Tr_T|g_-\rangle\langle g_-|
\right)^2
\right]
\nonumber\\
=&q_-.
\label{eq:supp-swapidentity}
\end{align}
The second equality is the standard two-copy SWAP identity for purity
~\cite{S-Ekert2002,S-Islam2015}. Related few-replica protocols provide alternative
routes to nonlinear quantum observables~\cite{S-LiuReplica2026}.

In a controlled-SWAP realization an ancilla is prepared in $|+\rangle$,
controls the exchange of the two image-address registers, and is measured in
the $X$ basis. Both work registers are measured to determine whether the
two copies lie in their good subspaces.

\subsection{Finite-shot estimator}

Let
$r=p_-^{(1)}p_-^{(2)}$
be the probability that both work registers are good in the ideal two-copy
experiment, and let $q$ denote the desired expectation. Define
\begin{equation}
Y=
\begin{cases}
+1,&\text{both copies good and SWAP outcome }+1,\\
-1,&\text{both copies good and SWAP outcome }-1,\\
0,&\text{otherwise}.
\end{cases}
\end{equation}
The exact probabilities are
\(P(Y=+1)=\frac{r+q}{2}, P(Y=-1)=\frac{r-q}{2}, P(Y=0)=1-r.\)
Thus
$\mathbb E[Y]=q,$
and for
$\hat q=
\frac{1}{N_s}\sum_{s=1}^{N_s}Y_s$
we have
\begin{equation}
\operatorname{Var}(\hat q)
=
\frac{r-q^2}{N_s}.
\label{eq:supp-qvar}
\end{equation}
This gives the ordinary $N_s^{-1/2}$ statistical scaling observed in the
numerical experiments.

\subsection{Destructive-SWAP readout}

The same observable can be measured without a controlled-SWAP ancilla.
Write the image-address SWAP as
$S_I=
\bigotimes_{\mu=1}^{b}S_\mu,$
where $b=\lceil\log_2B\rceil$ and $S_\mu$ exchanges the $\mu$th address qubit
between the two copies.

For each pair of corresponding address qubits, a Bell-basis measurement
resolves the symmetric and antisymmetric subspaces. Assign
\begin{equation}
s_\mu=
\begin{cases}
-1,&\text{antisymmetric Bell outcome },\\
+1,&\text{symmetric Bell outcome}.
\end{cases}
\end{equation}
The eigenvalue of the full address SWAP is then
$s_I=\prod_{\mu=1}^{b}s_\mu.$
A destructive single-shot estimator is therefore
\(Y_{\rm dest} = \mathbf 1_{\mathrm{good},1} \mathbf 1_{\mathrm{good},2} \,s_I.\)
Its expectation is again
$\mathbb E[Y_{\rm dest}]=q_-.$
A Bell-basis analysis can be implemented by one CNOT and one Hadamard per
address-qubit pair followed by computational-basis measurement. This removes
the controlled-SWAP ancilla and reduces two-qubit depth at the expense of
reading out more address qubits.

\subsection{Logical resources and implementation scope}

For batch size $B$, define
$b=\lceil\log_2B\rceil.$
One coherent overlap copy uses $2b+n$ logical qubits. Table~\ref{tab:supp-resource}
summarizes the resources that follow directly from the logical construction.

\begin{table}[h]
\caption{
Logical resources of the one- and two-copy observables.
Indexed loading and encoder decompositions are implementation dependent and
are therefore listed by coherent calls rather than native-gate count.
}
\label{tab:supp-resource}
\begin{ruledtabular}
\begin{tabular}{lcccc}
Observable
& Logical qubits
& Mask-loader calls
& Indexed encoder calls
& Additional operations
\\
\hline
$p_-$
& $2b+n$
& $1$
& $2$
& work-register measurement
\\
$q_-$, controlled SWAP
& $4b+2n+1$
& $2$
& $4$
& $b$ Fredkin gates
\\
$q_-$, destructive SWAP
& $4b+2n$
& $2$
& $4$
& $b$ Bell-pair analyses
\end{tabular}
\end{ruledtabular}
\end{table}

Here one ``indexed encoder call'' denotes one coherent multiplexed application
of either $U_j^T$ or $(U_i^I)^\dagger$. For the $B=4$, $n=2$ proof of principle, $b=2$. The controlled-SWAP circuit
therefore uses 13 logical qubits. The destructive readout requires two CNOTs
for its two Bell-pair analyses. The controlled implementation uses two
Fredkin gates; under the decomposition adopted in the logical-noise study,
this corresponds to 16 CNOTs. These counts refer only to the readout layer
and not to coherent indexed state preparation.

\subsection{Gradient of the two-copy observable}

Write the two-copy quantity with independently parameterized copies as
\(q(\theta_a,\theta_b) = \langle\Psi(\theta_a)| \langle\Psi(\theta_b)| O_q |\Psi(\theta_a)\rangle |\Psi(\theta_b)\rangle,\)
where
$O_q=
\Pi_0^{(1)}\Pi_0^{(2)}S_I.$
The physical objective is
$q_-(\theta)=q(\theta,\theta).$
For a Pauli-generated rotation with parameter-shift
$s=\pi/2$,
\(\partial_{\theta_a} q(\theta_a,\theta_b) \big|_{\theta_a=\theta_b=\theta}=\frac12\left[q(\theta+s,\theta)-q(\theta-s,\theta)\right].\)
Copy-exchange symmetry gives an identical contribution from the second copy.
Therefore
\begin{equation}
\frac{dq_-(\theta)}{d\theta}
=
q(\theta+s,\theta)
-
q(\theta-s,\theta).
\label{eq:supp-qshift}
\end{equation}
The one-copy probabilities $p_+=\frac{1}{B}\sum_iF_{ii}$ and $p_-$ obey the standard factor-$1/2$
parameter-shift rule. The complete loss gradient is thus obtained from
physically measurable shifted circuits.

\section{Quantum-native multistate representation-learning benchmark}

\subsection{Parameterized two-qubit family}

The transition-probability-indistinguishable family of
Eq.~\eqref{eq:supp-H4} is generated by
\begin{equation}
U(\varphi)
=
\operatorname{SWAP}
(I\otimes H)
\operatorname{CP}(\varphi)
(H\otimes I),
\label{eq:supp-qnative-U}
\end{equation}
where
$\operatorname{CP}(\varphi)
=
\operatorname{diag}(1,1,1,e^{i\varphi}).$
Up to the fixed basis-order convention, its columns equal
those of $H_4(\varphi)$. The coherent moment is
\begin{equation}
q_H(\varphi)
=
c+d\cos\varphi,
c=\frac{17}{576},
d=\frac{1}{144},
\label{eq:supp-qnative-harmonic}
\end{equation}
which depends on $\varphi$, contrary to the two-state transition probabilities.

Eight training geometries are
$\varphi_k
=
-\pi+\frac{2\pi k}{8},
k=0,\ldots,7,$
and the held-out phases are
$\varphi_k^{\rm test}
=
\varphi_k+\frac{\pi}{8}.$
Each positive view receives an independent perturbation
$\eta\sim\mathcal N(0,0.05^2)$ so that the training tests more than memorization, but generalization to interleaved relational geometries.

\subsection{Learned coherent probes}

A scalar $q_H$ cannot distinguish $\varphi$ from $-\varphi$. Define
\begin{equation}
\tilde{\bm z}_\delta(\varphi)=
\begin{pmatrix}
[q_H(\varphi)-c]/d\\
[q_H(\varphi+\delta)-c]/d
\end{pmatrix},
\bm z_\delta
=
\frac{\tilde{\bm z}_\delta}
{\|\tilde{\bm z}_\delta\|_2}.
\label{eq:supp-qnative-z}
\end{equation}
For independently augmented batches
$\{\bm z_i^a\}$ and $\{\bm z_i^b\}$, we optimize
\begin{equation}
\mathcal L_{\rm NCE}
=
-\frac{1}{16}\sum_{i=1}^{8}
\Bigg[
\log
\frac{
e^{\bm z_i^a\cdot\bm z_i^b/\tau}
}{
\sum_j
e^{\bm z_i^a\cdot\bm z_j^b/\tau}
}
+
\log
\frac{
e^{\bm z_i^b\cdot\bm z_i^a/\tau}
}{
\sum_j
e^{\bm z_i^b\cdot\bm z_j^a/\tau}
}
\Bigg],
\label{eq:supp-qnative-nce}
\end{equation}
with temperature hyperparemeter $\tau=0.2.$
Eq.~\eqref{eq:supp-qnative-harmonic} implies that
$\delta=\pm\pi/2$ supplies complementary cosine and sine quadratures and the
exact contrastive landscape has its minima at these values.

All eight training geometries are used in every epoch. The two Gaussian
perturbations defining each positive pair are resampled during training.
The only trainable parameter is $\delta$, initialized uniformly on
$[-\pi,\pi)$. If an initialization satisfies
$|\sin\delta|<0.05,$
it is displaced by $0.1$ rad to avoid beginning exactly at a symmetry-induced
zero-gradient point. This adjustment does not encode the quadrature solution.
We use 30 optimization seeds. A run is counted as learning the quadrature if
its final $\delta$ is within $0.25$ rad of either $+\pi/2$ or $-\pi/2$,
modulo $2\pi$. This criterion summarizes optimization success and is not used
in the loss.
For finite-shot training, the derivative of the second probe is obtained
from the one-copy-shifted two-copy observable,
\(\partial_\delta q_H(\varphi+\delta)=Q(\varphi+\delta+\pi/2,\varphi+\delta)-Q(\varphi+\delta-\pi/2,\varphi+\delta),\)
with the quoted shot count used independently for each shifted circuit.
Held-out retrieval is averaged over 100 independently perturbed test episodes
per optimization seed. Similarity is the dot product between normalized
two-component representations. We report symmetric R@1 and MRR in the table below.

\begin{table}[h]
\caption{
Quantum-native held-out multistate retrieval.
``Quad.'' gives the number of optimization runs that learn the quadrature.
Finite-shot R@1 values are mean $\pm$ standard deviation across optimization
seeds.
MRR is reported as mean only.
}
\label{tab:supp-qnative}
\begin{ruledtabular}
\begin{tabular}{lrrrr}
Noise & Shots & Quad. & R@1 & MRR (mean)\\
\hline
Ideal
& exact
& $30/30$
& $1.000$
& $1.000$
\\
Ideal
& 8192
& $28/30$
& $0.512\pm0.018$
& $0.720$
\\
Ideal
& 32768
& $30/30$
& $0.875\pm0.012$
& $0.937$
\\
Moderate
& exact
& $30/30$
& $1.000$
& $1.000$
\\
Moderate
& 8192
& $29/30$
& $0.443\pm0.019$
& $0.667$
\\
Moderate
& 32768
& $30/30$
& $0.812\pm0.014$
& $0.903$
\\
Harsh
& exact
& $29/30$
& $1.000$
& $1.000$
\\
Harsh
& 8192
& $19/30$
& $0.348\pm0.022$
& $0.585$
\\
Harsh
& 32768
& $30/30$
& $0.699\pm0.013$
& $0.840$
\end{tabular}
\end{ruledtabular}
\end{table}

A fidelity-only representation is identical for all eight geometries and
therefore has
$\mathrm{R@1}=0.125.$
A scalar single-$q_H$ representation reaches R@1 $=0.499$ in the ideal exact
setting, consistent with the residual
$\varphi\leftrightarrow-\varphi$ ambiguity. Under moderate noise it reaches
$0.251$ and $0.389$ at 8192 and 32768 shots, respectively, compared with
$0.443$ and $0.812$ for the learned two-probe representation.

\subsection{Logical noise and purity-readout study}
Multi-copy learning primitives can be sensitive to physical noise, and
recent theory emphasizes that noise can qualitatively change conclusions
associated with coherent quantum access~\cite{S-Cotler2026}. Here, we focus more on a specific study: to test whether the multistate observable remains a usable variational signal.

We use a logical gate-level error model for simplicity. The parameterized encoder of
Eq.~\eqref{eq:supp-qnative-U} is followed by a two-qubit Pauli channel with
total nonidentity probability $p_{\rm enc}$, distributed uniformly over the
15 nonidentity two-qubit Paulis. Each CNOT in the purity readout is followed
by the same channel with total error probability $p_2$. Every measured bit
is independently flipped with probability $e_r$. The work-register
measurements defining the good-space events are included, so false positive
and false negative good flags are explicitly modeled.
We have three stress settings $(p_{\rm enc},p_2,e_r)$: mild $(0.1\%,0.1\%,1\%)$, moderate (0.5\%,0.5\%,2\%) and harsh $(1\%,1\%,5\%)$.

For the endpoint geometries,
$\Delta q
=
q_H(0)-q_H(\pi)
=
\frac{1}{72}
=
0.013889.$
Under moderate noise the remaining separation is $0.01204$ for destructive
SWAP and $0.01224$ for controlled SWAP. Under harsh noise the corresponding
values are $0.00975$ and $0.01053$. A midpoint classifier distinguishes the endpoint geometries with
$86.6\%$ (destructive) and $86.8\%$ (controlled) accuracy at 2048 shots
under moderate noise, increasing to $98.6\%$ and $98.7\%$ at 8192 shots.
Under harsh noise the 8192-shot accuracies are $96.2\%$ and $97.2\%$.
For direct one-parameter minimization of $q_H$, we use 200 matched random
initializations, 100 Adam epochs, and an exponentially decaying learning rate
$0.3\rightarrow0.03$. Success is defined using the final iterate and 8192
shots per shifted-circuit expectation value. Under moderate noise,
$99.0\%$ of destructive-SWAP runs and $98.5\%$ of controlled-SWAP runs end
within $0.25$ rad of the optimum. Under harsh noise the corresponding
fractions are $88.5\%$ and $90.0\%$.

A fixed coherent phase offset
$\varphi_{\rm physical}
=
\varphi_{\rm control}+\delta_0$
primarily shifts the variational optimum and can therefore be absorbed by
training. We do not interpret this as strong robustness to coherent error and time-dependent stochastic drift would be a more stringent test.

\section{Microscopic topological-order benchmark}

\subsection{Loop-gas ground states and exact separation}

We construct explicit fixed-point toric-code and double-semion wavefunctions on periodic honeycomb lattices with qubits on edges~\cite{S-Kitaev2003,S-LevinWen2005,S-Zhang2020Signs}. An $L_x\times L_y$ torus contains
$N_q=3L_xL_y$
physical edge qubits. Closed-string configurations satisfy even parity at every trivalent vertex and separate into four homology sectors
$w=(w_x,w_y)\in\mathbb Z_2\times\mathbb Z_2.$
For each sector,
\begin{align}
|\Omega_w^{\rm TC}\rangle
&=
\frac{1}{\sqrt{N_w}}
\sum_{c\in\mathcal C_w}|c\rangle,
\\
|\Omega_w^{\rm DS}\rangle
&=
\frac{1}{\sqrt{N_w}}
\sum_{c\in\mathcal C_w}
(-1)^{N_\ell(c)}|c\rangle,
\label{eq:supp-topo-loopgas}
\end{align}
where $N_\ell(c)$ counts the closed-loop components of configuration $c$.

Dense state vectors are constructed explicitly for
$(L_x,L_y)=(2,2)/(3,2),$
corresponding to 12/18 physical qubits and 32/128 closed-string configurations, respectively. Sparse cycle-space representations are used for the 24- and 27-qubit size checks.
A minimally entangled ground-space basis is constructed by Fourier transforming one winding bit of the four homology-sector states. In this basis we use the canonical modular matrices
\(S_{\rm TC} = \frac12 \begin{pmatrix} 1&1&1&1\\ 1&1&-1&-1\\ 1&-1&1&-1\\ 1&-1&-1&1 \end{pmatrix},  T_{\rm TC} = \operatorname{diag}(1,1,1,-1),\)
and
$
S_{\rm DS}
=
\frac12
\begin{pmatrix}
1&1&1&1\\
1&-1&1&-1\\
1&1&-1&-1\\
1&-1&-1&1
\end{pmatrix},
T_{\rm DS}
=
\operatorname{diag}(1,i,-i,1).
$
These matrices encode the distinct anyonic statistics of the two topological orders~\cite{S-Smith2020Sign,S-Cian2022}.

Let $\{|a\rangle\}_{a=0}^{3}$ denote the logical ground-space basis. For each topological order we construct
\(|\psi_0\rangle=|0\rangle, |\phi_0\rangle=S|0\rangle, |\psi_1\rangle=T|1\rangle, |\phi_1\rangle=TS|0\rangle,\)
and lift these four logical rays into the corresponding microscopic ground space.
For both 12- and 18-qubit realizations, the complete six pairwise fidelities among the four microscopic rays are
\(\{F_{ab}\}_{\rm TC} = \{F_{ab}\}_{\rm DS} = \left\{ \frac14,0,\frac14,\frac14,\frac14,\frac14 \right\}.\)

Writing the four cross branches as a $2\times2$ overlap matrix $A$, the same construction as in Eq.~\eqref{eq:supp-sigma} gives
$\Sigma=\frac14AA^\dagger.$
Both topological orders have
$p=\Tr\Sigma=\frac14,$
but
\begin{equation}
q_{\rm TC}
=
\Tr\Sigma_{\rm TC}^2
=
\frac1{16},
q_{\rm DS}
=
\Tr\Sigma_{\rm DS}^2
=
\frac3{64}.
\label{eq:supp-topo-q}
\end{equation}
The corresponding Bargmann invariant has
\(\arg\mathcal B_{\rm TC}=0,  \arg\mathcal B_{\rm DS}=\frac{\pi}{2}.\)
Thus the microscopic embedding exactly preserves the fidelity-blind topological separation.

The loop-gas wavefunctions, homology sectors, and all local physical-qubit perturbations in this benchmark are constructed explicitly. The modular $S$ and $T$ transformations defined above are implemented as exact logical transformations lifted into the microscopic topological ground space. We do not decompose them into a local Dehn-twist circuit or an explicit sequence of microscopic Wilson-loop operations. Such operator-level constructions are known routes to extracting braiding statistics and topological twists~\cite{S-Cian2022}, but are outside the scope of the present benchmark.

\subsection{Local physical-qubit perturbations}

Each of the four prepared many-body rays is independently perturbed on every physical edge by
\(U_a = \bigotimes_{e=1}^{N_q} R_x(\theta_{a,e}),  \theta_{a,e}\sim\mathcal N(0,\sigma^2),\)
with independent angles for every ray $a$ and edge $e$. These rotations mix distinct closed-loop configurations and therefore perturb the microscopic wavefunction.

For each topological order and each $\sigma$, 500 independently perturbed relational quartets are generated on the 12- and 18-qubit systems. Classification uses Euclidean nearest-centroid decision rules. In each of 300 independent episodes, five examples from each topological order are selected as calibration examples and the remaining examples are used for evaluation.
The pairwise baseline receives the complete six-dimensional vector
$(F_{01},F_{02},F_{03},F_{12},F_{13},F_{23}),$
whereas the coherent classifier receives only the scalar $q$.

\begin{table}[h]
\caption{
Topological-order identification on the 18-qubit microscopic loop-gas
benchmark under independent local $R_x$ perturbations on every physical
qubit. Values are mean $\pm$ standard deviation over 300 independent calibration episodes.
}
\label{tab:supp-topo-robustness}
\begin{ruledtabular}
\begin{tabular}{ccc}
$\sigma$
& Six pairwise fidelities
& Coherent $q$
\\
\hline
0.00
& $0.5000\pm0.0000$
& $1.0000\pm0.0000$
\\
0.02
& $0.4978\pm0.0116$
& $1.0000\pm0.0000$
\\
0.05
& $0.5034\pm0.0215$
& $1.0000\pm0.0000$
\\
0.10
& $0.4994\pm0.0160$
& $0.9986\pm0.0015$
\\
0.15
& $0.4999\pm0.0153$
& $0.9581\pm0.0099$
\\
0.20
& $0.4997\pm0.0151$
& $0.8180\pm0.0178$
\\
0.30
& $0.5053\pm0.0218$
& $0.6649\pm0.0898$
\end{tabular}
\end{ruledtabular}
\end{table}

\subsection{System-size check}

To reach larger microscopic systems without forming dense $2^{N_q}$ vectors, we enumerate the closed-string cycle space and store the loop-gas states sparsely. Each prepared ray receives two independent local $R_x$ defects with rotation angles drawn from
$\mathcal N(0,0.2^2).$
Results using five calibration examples per class are summarized in the table below, which shows that he fidelity-blind distinction persists across different microscopic system sizes.

\begin{table}[h]
\caption{
Topological-order identification versus microscopic system size for two local
$R_x$ defects per prepared ray.
}
\label{tab:supp-topo-size}
\begin{ruledtabular}
\begin{tabular}{ccc}
Physical qubits
& Six pairwise fidelities
& Coherent $q$
\\
\hline
12
& $0.4989\pm0.0146$
& $0.9938\pm0.0038$
\\
18
& $0.4990\pm0.0123$
& $0.9964\pm0.0013$
\\
24
& $0.4979\pm0.0218$
& $0.9918\pm0.0033$
\\
27
& $0.5113\pm0.0382$
& $0.9837\pm0.0080$
\end{tabular}
\end{ruledtabular}
\end{table}

\subsection{Finite-shot readout}

Finite-shot values of $q$ are sampled using the two-copy estimator. For the pairwise baseline, every one of the six fidelities is estimated independently with the same number of shots, giving the pairwise method a larger total measurement budget.
The same 27-qubit system studied above is evaluated on 200 independent episodes. Each episode uses five calibration measurements per phase and 50 test examples per phase.

\begin{table}[h]
\caption{
Finite-shot 27-qubit topological-order identification.
The quoted pairwise shot count is used independently for each of the six
measured fidelities.
}
\label{tab:supp-topo-shots}
\begin{ruledtabular}
\begin{tabular}{rcc}
Shots per observable
& Six pairwise fidelities
& Coherent $q$
\\
\hline
1024
& $0.4980\pm0.0465$
& $0.8026\pm0.0390$
\\
2048
& $0.5038\pm0.0484$
& $0.8794\pm0.0343$
\\
4096
& $0.5100\pm0.0501$
& $0.9372\pm0.0264$
\\
8192
& $0.5106\pm0.0514$
& $0.9689\pm0.0170$
\end{tabular}
\end{ruledtabular}
\end{table}

\section{Four-photon collective-phase learning benchmark}

This section gives the protocol for the four-photon collective-phase benchmark.
The construction is parameterized by the experiment of
Jones \emph{et al.}~\cite{S-Jones2020}, and all samples used here are generated
from the model below.  The
learner must recover a continuously varying collective phase from relational
measurements while source parameters drift independently between views.  Phase
labels are not supplied during representation learning.

\subsection{Four-photon internal states and collective phase}

We consider four single-photon internal states, denoted
$|a\rangle,|b\rangle,|c\rangle,|d\rangle$.  Polarization and temporal degrees
of freedom are written explicitly as
\begin{align}
|a\rangle
&=|H\rangle|\tau_1\rangle,
&
|c\rangle
&=|V\rangle|\tau_1\rangle,
\\
|b\rangle
&=
(\cos\beta_b|H\rangle+\sin\beta_b|V\rangle)|\tau_2\rangle,
&
|d\rangle
&=
(\cos\beta_d|H\rangle+e^{i\varphi}\sin\beta_d|V\rangle)
|\tau_3\rangle .
\label{eq:supp-photon-states}
\end{align}
The latent variable $\varphi\in[0,2\pi)$ is the collective phase to be
learned.  We take the temporal overlaps
\(x_{12}=\langle\tau_1|\tau_2\rangle,
x_{13}=\langle\tau_1|\tau_3\rangle,
x_{23}=\langle\tau_2|\tau_3\rangle\)
to be real.  In the ideal pairwise-distinguishable limit $x_{23}=0$, the
phase is absent from every pairwise fidelity even though it remains in a
four-state interference loop.

In the order $(a,b,c,d)$, the undamped Gram matrix
$G^{(0)}_{ij}=\langle i|j\rangle$ is determined by
\begin{align}
G^{(0)}_{ab}&=x_{12}\cos\beta_b,
&G^{(0)}_{ac}&=0,
&G^{(0)}_{ad}&=x_{13}\cos\beta_d,
\\
G^{(0)}_{bc}&=x_{12}\sin\beta_b,
&G^{(0)}_{cd}&=x_{13}\sin\beta_d e^{i\varphi},
\label{eq:supp-photon-gram-elements}
\\
G^{(0)}_{bd}
&=x_{23}
\left(
\cos\beta_b\cos\beta_d
+\sin\beta_b\sin\beta_d e^{i\varphi}
\right),
\nonumber
\end{align}
with $G^{(0)}_{ji}=G^{(0)*}_{ij}$ and unit diagonal.  Residual spectral
mismatch is represented by a fidelity parameter $\eta$ through
\begin{equation}
G=\sqrt{\eta}\,G^{(0)}+(1-\sqrt{\eta})I.
\label{eq:supp-photon-spectral-damping}
\end{equation}
Thus every off-diagonal amplitude is multiplied by $\sqrt\eta$, while the
diagonal remains one.  Equation~\eqref{eq:supp-photon-spectral-damping} is a
convex combination of Gram matrices and therefore preserves positivity once
the temporal overlaps themselves define a physical correlation matrix.  This
requires
\begin{equation}
\det T
=
1+2x_{12}x_{13}x_{23}
-x_{12}^2-x_{13}^2-x_{23}^2
\geq0,
\label{eq:supp-photon-temporal-psd}
\end{equation}
in addition to the $2\times2$ principal-minor conditions.  An audit of one
million independent box proposals found that $3.63\%$ of the nominal
training or in-distribution (ID) temporal triples violated
Eq.~\eqref{eq:supp-photon-temporal-psd} while the OOD test data was entirely physical.
All reported data therefore use rejection sampling from the stated boxes,
conditioned only on temporal positive semidefiniteness.  In a fresh audit of
$10^5$ accepted samples per split, the minimum temporal/full-Gram eigenvalues
were respectively $5.99\times10^{-6}/9.53\times10^{-4}$ (training),
$5.56\times10^{-6}/8.41\times10^{-4}$ (ID), and
$7.39\times10^{-2}/9.38\times10^{-2}$ (OOD).  Thus every retained example
admits an explicit temporal-mode realization.

For these four states we use the standard paired negative mask
$\mathcal N=\{(i,j):i\neq j\}$, with $B=4$ and $M=12$, and define
$A_-$, $\Sigma_-$, $p_-$, and $q_-$ exactly as in
Eqs.~\eqref{eq:qminus} and \eqref{eq:supp-sigma}.  Equivalently,
\begin{align}
p_-(\varphi)
=
\frac{1}{12}
\sum_{i\neq j}|G_{ij}(\varphi)|^2,
\quad q_-(\varphi)
=
\frac{1}{144}
\Tr\!\left[
(A_-A_-^\dagger)^2
\right].
\label{eq:supp-photon-pq}
\end{align}
Because the overall overlap scale also drifts, the learning feature is the
conditional purity
\begin{equation}
\bar q_-(\varphi)
\equiv
\frac{q_-(\varphi)}{p_-^2(\varphi)}.
\label{eq:supp-photon-qbar}
\end{equation}
The quantity $p_-$ is the mean of the six unique pairwise fidelities, so
Eq.~\eqref{eq:supp-photon-qbar} requires no additional measurement setting.
The normalization removes changes in total negative weight and retains the
spectral shape of $\Sigma_-$.

At $x_{23}=0$, all six pairwise fidelities are independent of $\varphi$.
A nonzero $x_{23}$ leaks a phase-dependent term into $F_{bd}$, but its
amplitude is multiplied by the unknown source variables $x_{23}$, $\beta_b$,
and $\beta_d$.  With two phase-shifted settings and independently drifting
nuisances, the inverse problem is ambiguous.  We therefore test $\bar q_-$
against the complete two-probe fidelity baseline, a fixed-$x_{23}$ control,
expanded three-probe fidelity baselines, and the protected $x_{23}=0$ limit.

\subsection{Synthetic data and pairwise baseline}

For every example, the phase and nuisance proposals are drawn independently,
after which nonphysical temporal triples are rejected.  The latent phase is
continuous,
$\varphi\sim\operatorname{Unif}[0,2\pi)$,
and the two views of a positive contrastive pair share only $\varphi$.
Their accepted temporal triples are independently realized between views, as
are their polarization angles, spectral mismatch, and measurement noise.  The
parameter ranges are listed in Table~\ref{tab:supp-photon-data}.

\begin{table}[h]
\caption{
Distributions used to generate the four-photon data.  Both $x_{12}$ and
$x_{13}$ are proposed independently from the interval shown.  The polarization
angles obey $\beta_b,\beta_d\sim\mathcal N(\pi/4,\sigma_\beta^2)$.  Box
proposals are mutually independent before conditioning the temporal triple on
Eq.~\eqref{eq:supp-photon-temporal-psd}.
}
\label{tab:supp-photon-data}
\begin{ruledtabular}
\begin{tabular}{lcccc}
Split
& $x_{12},x_{13}$
& $x_{23}$
& $\sigma_\beta$
& $\eta$
\\
\hline
Training / ID
& $[0.68,0.72]$
& $[0,0.07]$
& $0.025$
& $[0.95,1]$
\\
Compound OOD
& $[0.63,0.68]$
& $[0.07,0.10]$
& $0.050$
& $[0.92,0.97]$
\\
Fixed-$x_{23}$ training
& $[0.68,0.72]$
& $0.10$
& $0.025$
& $[0.95,1]$
\\
Fixed-$x_{23}$ OOD
& $[0.63,0.68]$
& $0.10$
& $0.050$
& $[0.92,0.97]$
\\
Ideal readout curve
& $1/\sqrt2$
& $0$
& $0$
& $1$
\end{tabular}
\end{ruledtabular}
\end{table}

Training and ID evaluation use the same parameter ranges but independently
generated samples.  Compound OOD evaluation simultaneously shifts all four
nuisance families.  The fixed-$x_{23}$ control retains the temporal,
polarization, and spectral shift but removes uncertainty in the overlap that
leaks the phase into $F_{bd}$.

Let $\bm f(\varphi)=(F_{ab},F_{ac},F_{ad},F_{bc},F_{bd},F_{cd})$
contain the complete set of six unique pairwise fidelities.  A second probe
applies the known control offset
$\delta=\frac{\pi}{2}$
to the phase of $|d\rangle$.  The feature families are
\begin{align}
\bm x_F^{(1)}(\varphi)
&=
\bm f(\varphi),
\\
\bm x_F^{(2)}(\varphi)
&=
\bigl[
\bm f(\varphi),
\bm f(\varphi+\delta)
\bigr],
\\
\bm x_{\rm coh}^{(1)}(\varphi)
&=
\bigl[
\bm f(\varphi),
\bar q_-(\varphi)
\bigr],
\\
\bm x_{\rm coh}^{(2)}(\varphi)
&=
\bigl[
\bm f(\varphi),
\bm f(\varphi+\delta),
\bar q_-(\varphi),
\bar q_-(\varphi+\delta)
\bigr].
\label{eq:supp-photon-features}
\end{align}
Their dimensions are $6$, $12$, $7$, and $14$, respectively.  The principal
baseline is $\bm x_F^{(2)}$: it receives every pairwise fidelity at both phase
quadratures.  The coherent learner receives the same 12 fidelity values and
adds only the two conditional purities.

For the shuffled-$q$ control, the two $\bar q_-$ coordinates are permuted
jointly across examples after measurement.  This preserves their marginal
distribution and the feature dimension while destroying their association
with the phase and the corresponding fidelity vector.

\subsection{Finite-shot estimators and label-free tied-CCA learning protocol}

For a pairwise fidelity with exact mean $F$, finite-shot simulations use the
Gaussian approximation to a Bernoulli estimator,
\[
\widehat F
=
F+\xi_F,
\qquad
\operatorname{Var}(\xi_F)
=
\frac{F(1-F)}{N_s},
\]
followed by clipping to $[0,1]$.  The two-copy estimator of
Eq.~\eqref{eq:supp-qvar} has outcomes $Y\in\{-1,0,+1\}$ with
\[
\mathbb E[Y]=q_-,
\qquad
\mathbb E[Y^2]=p_-^2,
\qquad
\operatorname{Var}(\widehat q_-)
=
\frac{p_-^2-q_-^2}{N_s}.
\]
The noisy conditional purity is formed as
$\widehat{\bar q}_-=\widehat q_-/\widehat p_-^2$, where
$\widehat p_-$ is computed from the six noisy fidelities.  Shot noise is
sampled independently in the two contrastive views.

The per-observable comparisons assign the same $N_s$ to every reported
scalar.  We additionally fix the total budget to $N_{\rm tot}=393216$.
The 12-scalar pairwise baseline then uses $32768$ shots per scalar, whereas
the 14-scalar coherent input uses
$\lfloor N_{\rm tot}/14\rfloor=28086$ shots per scalar.  Thus the
equal-budget test does not grant the coherent model extra measurements.

The primary representation learner is a tied linear canonical correlation analysis (CCA) model, equivalently the
closed-form linear limit of a redundancy-reduction contrastive objective.
For each of ten seeds, we generate $N=50000$ positive pairs.  The two views in
a pair share a uniformly sampled continuous phase but have independent
nuisance parameters drawn from the training distribution.

Let $X^{(1)},X^{(2)}\in\mathbb R^{N\times d}$ be the two standardized feature
matrices.  We form the symmetrized within-view and cross-view covariances
\begin{equation}
C
=
\frac{
X^{(1)\mathsf T}X^{(1)}
+
X^{(2)\mathsf T}X^{(2)}
}{2N}
+
\gamma I, \quad
C_\times
=
\frac{
X^{(1)\mathsf T}X^{(2)}
+
X^{(2)\mathsf T}X^{(1)}
}{2N},
\label{eq:supp-photon-cca-cov}
\end{equation}
with $\gamma=10^{-3}$.  If $V_2$ contains the two leading eigenvectors of
$C^{-1/2}C_\times C^{-1/2}$, the tied encoder is
\begin{equation}
\bm z(\bm x)
=
(\bm x-\bm\mu)
D^{-1}
C^{-1/2}
V_2,
\label{eq:supp-photon-cca-encoder}
\end{equation}
where $\bm\mu$ and the diagonal scale $D$ are obtained from the pooled views.
Neither $\varphi$ nor any supervised target enters
Eqs.~\eqref{eq:supp-photon-cca-cov}--\eqref{eq:supp-photon-cca-encoder}.

Evaluation uses 16 calibration phases $\varphi_k=\frac{2\pi k}{16},k=0,\ldots,15$, and 16 interleaved held-out phases $\varphi'_k = \frac{2\pi(k+1/2)}{16}$.
For each phase, 256 independently drifted views are averaged to form a
prototype.  A linear map from the 16 calibration prototypes to
$(\cos\varphi_k,\sin\varphi_k)$ is fit after the encoder is frozen.  The phase
metric is the circular mean absolute error (MAE)
\[
\operatorname{MAE}_{\rm ang}
=
\frac1{16}
\sum_k
\left|
\arg e^{i(\widehat\varphi'_k-\varphi'_k)}
\right|.
\]
Retrieval R@1 uses independent OOD queries and assigns each query to the
nearest held-out phase prototype.  We also report the second canonical
correlation $\rho_2$, since two independent quadratures are required to embed
a circle.

\subsection{Primary contrastive results}

Tab.~\ref{tab:supp-photon-cca-ablation} shows the exact-expectation feature
ablation.  A single probe does not determine the orientation of the phase
circle.  Two complete fidelity probes improve ID interpolation but fail under
compound OOD drift.  Adding both coherent quadratures produces an essentially
perfect two-dimensional phase representation.  Shuffling $\bar q_-$ restores
the fidelity-only result.

\begin{table}[h]
\caption{
Exact-expectation tied-CCA feature ablation under compound OOD drift.  Values
are mean $\pm$ standard deviation over ten seeds.  The last column is the
second canonical correlation learned without labels.
}
\label{tab:supp-photon-cca-ablation}
\begin{ruledtabular}
\begin{tabular}{lccc}
Features
& Angular MAE (rad)
& R@1
& $\rho_2$
\\
\hline
$\bm x_F^{(1)}$
& $1.342 \pm 0.067$
& $0.138 \pm 0.021$
& $0.0081 \pm 0.0032$
\\
$\bm x_F^{(2)}$
& $0.940 \pm 0.020$
& $0.645 \pm 0.024$
& $0.2094 \pm 0.0068$
\\
$\bm x_{\rm coh}^{(1)}$
& $1.296 \pm 0.183$
& $0.116 \pm 0.043$
& $0.0116 \pm 0.0015$
\\
$\bm x_{\rm coh}^{(2)}$
& $0.00503 \pm 0.00021$
& $1.0000 \pm 0.0000$
& $0.998965 \pm 0.000004$
\\
$\bm x_{\rm coh}^{(2)}$, shuffled $\bar q_-$
& $0.967 \pm 0.020$
& $0.646 \pm 0.024$
& $0.2130 \pm 0.0043$
\end{tabular}
\end{ruledtabular}
\end{table}

On the ID split, the corresponding two-probe angular errors are
$0.0440\pm0.0115$ for $\bm x_F^{(2)}$,
$0.000574\pm0.000051$ for $\bm x_{\rm coh}^{(2)}$,
and $0.0437\pm0.0114$ after shuffling $\bar q_-$.  Thus the all-fidelity learner
can interpolate within the training distribution while its failure in
Tab.~\ref{tab:supp-photon-cca-ablation} is specifically a failure to identify
the phase under compound nuisance shift.

The same comparison remains separated at finite shots and under the gate
model, as summarized in Tab.~\ref{tab:supp-photon-cca-noise}.  In every
condition, the shuffled control tracks the all-fidelity baseline.  At fixed
total budget, the coherent method uses fewer shots per scalar yet reduces the
mean angular error from $0.939\pm0.022$ to $0.131\pm0.005$ rad.

\begin{table}[h]
\caption{
Tied-CCA generalization under compound OOD drift.  All entries are angular
MAE in radians over ten seeds.  ``Moderate gates'' includes $32768$ shots and
the readout visibility obtained from the gate-level audit below.
}
\label{tab:supp-photon-cca-noise}
\begin{ruledtabular}
\begin{tabular}{lccc}
Condition
& $\bm x_F^{(2)}$
& $\bm x_{\rm coh}^{(2)}$
& Shuffled $\bar q_-$
\\
\hline
Exact
& $0.940 \pm 0.020$
& $0.00503 \pm 0.00021$
& $0.967 \pm 0.020$
\\
$8192$ shots per scalar
& $0.931 \pm 0.022$
& $0.245 \pm 0.008$
& $0.959 \pm 0.026$
\\
$32768$ shots per scalar
& $0.939 \pm 0.018$
& $0.118 \pm 0.005$
& $0.970 \pm 0.019$
\\
Moderate gates
& $0.939 \pm 0.020$
& $0.142 \pm 0.004$
& $0.962 \pm 0.020$
\\
Fixed $N_{\rm tot}=393216$
& $0.939 \pm 0.022$
& $0.131 \pm 0.005$
& $0.966 \pm 0.021$
\end{tabular}
\end{ruledtabular}
\end{table}

\subsection{Probe-number ablation and measurement-interface boundary}

The main comparison fixes two phase settings, $0$ and $\pi/2$.  To determine
whether the separation survives an expanded pairwise measurement interface,
we also evaluate nested fidelity probes at $(0,\pi/2,\pi)$ and balanced probes
at $(0,2\pi/3,4\pi/3)$.  All six fidelities are measured at every setting.
The balanced design isolates the two phase quadratures from the unknown DC
component by a well-conditioned discrete Fourier transform and is the
strongest three-probe baseline considered here.

\begin{table}[h]
\caption{
Tied-CCA angular MAE under drifting leakage for the probe-number ablation.
Values are mean $\pm$ standard deviation over ten seeds.  The fixed-budget
row uses $N_{\rm tot}=393216$ shots.  Subscripts $2$, $3n$, and $3b$ denote
two, three nested, and three balanced probe settings.
}
\label{tab:supp-photon-three-probe}
\begin{ruledtabular}
\begin{tabular}{lcccc}
Condition
& $F_2$
& $F_{3n}$
& $F_{3b}$
& $(F+\bar q_-)_{2}$
\\
\hline
Exact
& $0.957 \pm 0.024$
& $0.0092 \pm 0.0038$
& $0.0090 \pm 0.0037$
& $0.00492 \pm 0.00018$
\\
$8192$ shots per scalar
& $0.968 \pm 0.019$
& $0.0496 \pm 0.0090$
& $0.0131 \pm 0.0017$
& $0.242 \pm 0.009$
\\
$32768$ shots per scalar
& $0.970 \pm 0.015$
& $0.0157 \pm 0.0060$
& $0.0107 \pm 0.0023$
& $0.117 \pm 0.005$
\\
Fixed total budget
& $0.970 \pm 0.015$
& $0.0209 \pm 0.0075$
& $0.0111 \pm 0.0021$
& $0.128 \pm 0.005$
\end{tabular}
\end{ruledtabular}
\end{table}

A third balanced pairwise setting therefore restores identifiability in the
generic drifting-leakage model.  This indicates that the main
advantage is relative to the two-probe measuring interface. The prototype-based
CCA evaluation averages 256 independently drifted views per phase, so the
finite-shot errors in Tab.~\ref{tab:supp-photon-three-probe} should not be
interpreted as single-shot estimator errors.

There is nevertheless a physically meaningful limit in which adding such
settings cannot help.  We set $x_{23}=0$ exactly while retaining the other
training and OOD drifts.  This represents strictly orthogonal temporal or
frequency bins, or a mode relation protected by a symmetry.  Then
$F_{bd}(\varphi+\delta)=0$ for every control phase $\delta$, and all remaining
pairwise fidelities are also phase independent.

\begin{table}[h]
\caption{
Tied-CCA angular MAE in the protected-orthogonality control.  Values are mean
$\pm$ standard deviation over ten seeds.  The random-phase value is $\pi/2$;
probe subscripts follow Table~\ref{tab:supp-photon-three-probe}.
}
\label{tab:supp-photon-protected}
\begin{ruledtabular}
\begin{tabular}{lcccc}
Condition
& $F_2$
& $F_{3n}$
& $F_{3b}$
& $(F+\bar q_-)_{2}$
\\
\hline
Exact
& $1.569 \pm 0.005$
& $1.568 \pm 0.005$
& $1.571 \pm 0.007$
& $0.00132 \pm 0.00016$
\\
$32768$ shots per scalar
& $1.570 \pm 0.019$
& $1.569 \pm 0.006$
& $1.569 \pm 0.006$
& $0.0225 \pm 0.0061$
\end{tabular}
\end{ruledtabular}
\end{table}

The pairwise representations remain at random-phase error while the coherent
moment retains the collective phase.  Exact orthogonality is not generic to
source drift, so experimental use of this stronger separation requires a
specific mode-engineering or symmetry mechanism that protects $x_{23}=0$.

\subsection{Independent supervised identifiability control}

As a model-class check, we generate 10000 continuous training phases and
3000 independently sampled test phases per seed and regress the targets
$(\cos\varphi,\sin\varphi)$.  We use an ExtraTrees regressor with 256 trees,
minimum leaf size two, and all input features available at each split.  Under compound OOD drift and exact
expectations, the two-probe pairwise input gives angular MAE $0.550\pm0.016$ rad,
whereas the coherent input gives $0.0207\pm0.0021$ rad. Again, shuffling $\bar q_-$ returns
the error to $0.552\pm0.015$ rad.

The fixed-$x_{23}$ control isolates the role of the drifting leakage term.
On the fixed-$x_{23}$ OOD split, the two-probe fidelity baseline reaches
$\operatorname{MAE}_{\rm ang}=0.0266\pm0.0013$ rad, compared with $0.0191\pm0.0011$ rad
for the coherent input and $0.0267\pm0.012$ rad after shuffling $\bar q_-$.  Thus
pairwise data solve the task once the relevant nuisance is known.  A
degree-two polynomial ridge model gives $0.00234\pm0.00013$ rad with the normalized
two-probe coherent input, confirming that the separation is not specific to
tree ensembles.

\subsection{Destructive-SWAP gate and readout-noise audit}

The gate-level audit starts after coherent indexed preparation of the two
copies in Eq.~\eqref{eq:q-circuit}.  Since $B=4$, the $I$ register contains
two address qubits per copy.  A destructive SWAP measurement Bell-analyzes
each corresponding address pair with one CNOT and one Hadamard, for a total of
$N_{\rm CNOT}=2$ and $N_H=2$, followed by measurement of four address qubits.
The shot is accepted only when both work registers are measured in the good
subspace.  The ideal decomposition reproduces $q_-$ with maximum absolute
error $3.47\times10^{-18}$.

After every one- and two-qubit gate we apply a uniform nonidentity Pauli
channel with total probabilities $\epsilon_1$ and $\epsilon_2$,
respectively.  Each address measurement has a symmetric flip probability
$r$, and each work qubit has false-negative probability $r_W$.  The channels
are propagated exactly in the Heisenberg picture.  The resulting response is
nearly affine as
$q_{\rm gate}=a q_-+b$.
Tab.~\ref{tab:supp-photon-gate-models} reports the noise parameters, fitted
slope, and the root mean squared error (RMSE) of the inverse-affine conditional-purity estimate at
32768 shots. This isolated readout test uses the exact $p_-$ in the
denominator while the learning experiments instead use $\widehat p_-$ from the
noisy fidelity measurements.

\begin{table}[h]
\caption{
Gate and readout-noise models.  The last column is the calibrated RMSE of
$\bar q_-$ at 32768 shots over 41 phases and 400 independent shot repetitions.
}
\label{tab:supp-photon-gate-models}
\begin{ruledtabular}
\begin{tabular}{lcccccc}
Model
& $\epsilon_1$
& $\epsilon_2$
& $r$
& $r_W$
& $a$
& RMSE$(\bar q_-)$
\\
\hline
Ideal
& $0$
& $0$
& $0$
& $0$
& $1.000$
& $0.0331$
\\
Mild
& $2\!\times\!10^{-4}$
& $0.004$
& $0.01$
& $0.01$
& $0.938$
& $0.0347$
\\
Moderate
& $0.001$
& $0.010$
& $0.02$
& $0.02$
& $0.876$
& $0.0360$
\\
Harsh
& $0.003$
& $0.030$
& $0.04$
& $0.04$
& $0.755$
& $0.0408$
\end{tabular}
\end{ruledtabular}
\end{table}

Across 256 compound-OOD states, the maximum residuals from an affine response
are $3.58\times10^{-5}$, $6.88\times10^{-5}$, and $1.26\times10^{-4}$ for the
mild, moderate, and harsh models.  For the learning-level ``moderate gates''
condition, the measured slope is rounded conservatively to a residual
visibility of $0.88$, 32768 shots are used per scalar, and the pairwise
baseline is granted noiseless gates.  The static attenuation is learnable
from training data. The reported OOD result therefore tests shot noise and distribution shift in addition to the reduced coherent signal.

\begingroup
\renewcommand{\bibsection}{\section*{Supplemental References}}
\begin{suppbibliography}{99}

\bibitem{S-Bargmann1964}
V. Bargmann,
Note on Wigner's Theorem on Symmetry Operations,
J. Math. Phys. \textbf{5}, 862 (1964).

\bibitem{S-Rabei1999}
E. M. Rabei, Arvind, N. Mukunda, and R. Simon,
Bargmann Invariants and Geometric Phases: A Generalized Connection,
Phys. Rev. A \textbf{60}, 3397 (1999).

\bibitem{S-Mukunda2003}
N. Mukunda, Arvind, E. Ercolessi, G. Marmo, G. Morandi, and R. Simon,
Bargmann Invariants, Null Phase Curves, and a Theory of the Geometric Phase,
Phys. Rev. A \textbf{67}, 042114 (2003).

\bibitem{S-Oszmaniec2024}
M. Oszmaniec, D. J. Brod, and E. F. Galv\~ao,
Measuring Relational Information between Quantum States, and Applications,
New J. Phys. \textbf{26}, 013053 (2024).

\bibitem{S-LiWagnerZhang2026}
M.-S. Li, R. Wagner, and L. Zhang,
Multistate Imaginarity and Coherence in Qubit Systems,
Phys. Rev. A \textbf{113}, 012428 (2026).

\bibitem{S-ChienWaldron2016}
T.-Y. Chien and S. Waldron,
A Characterization of Projective Unitary Equivalence of Finite Frames and Applications,
SIAM J. Discrete Math. \textbf{30}, 976 (2016).

\bibitem{S-Tadej2006}
W. Tadej and K. \.{Z}yczkowski,
A Concise Guide to Complex Hadamard Matrices,
Open Syst. Inf. Dyn. \textbf{13}, 133 (2006).

\bibitem{S-Ekert2002}
A. K. Ekert, C. M. Alves, D. K. L. Oi, M. Horodecki, P. Horodecki, and L. C. Kwek,
Direct Estimations of Linear and Nonlinear Functionals of a Quantum State,
Phys. Rev. Lett. \textbf{88}, 217901 (2002).

\bibitem{S-Islam2015}
R. Islam, R. Ma, P. M. Preiss, M. E. Tai, A. Lukin, M. Rispoli, and M. Greiner,
Measuring Entanglement Entropy in a Quantum Many-Body System,
Nature \textbf{528}, 77 (2015).

\bibitem{S-LiuReplica2026}
Q. Liu, Z. Li, X. Yuan, H. Zhu, and Y. Zhou,
Auxiliary-Free Replica Shadows: Efficient Estimation of Multiple Nonlinear Quantum Properties,
Phys. Rev. Lett. \textbf{136}, 100602 (2026).

\bibitem{S-Cotler2026}
J. Cotler, W. Gong, and I. Kannan,
Noisy Quantum Learning Theory,
Nat. Commun. \textbf{17}, 6979 (2026).

\bibitem{S-Kitaev2003}
A. Yu. Kitaev,
Fault-Tolerant Quantum Computation by Anyons,
Ann. Phys. \textbf{303}, 2 (2003).

\bibitem{S-LevinWen2005}
M. A. Levin and X.-G. Wen,
String-Net Condensation: A Physical Mechanism for Topological Phases,
Phys. Rev. B \textbf{71}, 045110 (2005).

\bibitem{S-Zhang2020Signs}
Q. Zhang, W.-T. Xu, Z.-Q. Wang, and G.-M. Zhang,
Non-Hermitian Effects of the Intrinsic Signs in Topologically Ordered Wavefunctions,
Commun. Phys. \textbf{3}, 209 (2020).

\bibitem{S-Smith2020Sign}
A. Smith, O. Golan, and Z. Ringel,
Intrinsic Sign Problems in Topological Quantum Field Theories,
Phys. Rev. Research \textbf{2}, 033515 (2020).

\bibitem{S-Cian2022}
Z.-P. Cian, M. Hafezi, and M. Barkeshli,
Extracting Wilson Loop Operators and Fractional Statistics from a Single Bulk Ground State,
arXiv:2209.14302 (2022).

\bibitem{S-Jones2020}
A. E. Jones, A. J. Menssen, H. M. Chrzanowski, T. A. W. Wolterink,
V. S. Shchesnovich, and I. A. Walmsley,
Multiparticle Interference of Pairwise Distinguishable Photons,
Phys. Rev. Lett. \textbf{125}, 123603 (2020).

\end{suppbibliography}
\endgroup

\end{document}